\documentclass[prb,twocolumn,floatfix]{revtex4}

\usepackage{amsfonts}
\usepackage{amsmath}
\usepackage{verbatim}
\usepackage[dvips]{graphicx}
\usepackage{subfigure}

\def\Tr{\textrm}
\def\dd{\textrm{d}}

\def\vv{\Tr{v}}
\def\pp{\Tr{p}}

\begin{document}

\title{Fractional topological insulators of Cooper pairs induced by proximity effect}
\author{Predrag Nikoli\'c$^{1,2}$, Tanja Duric$^{1}$ and Zlatko Te\v{s}anovi\'{c}$^{2}$}
\affiliation{$^1$School of Physics, Astronomy and Computational Sciences,\\George Mason University, Fairfax, VA 22030, USA}
\affiliation{$^2$Institute for Quantum Matter, Johns Hopkins University, Baltimore, MD 21218, USA}
\date{\today}

% PACS:
% 73.90.+f	Other topics in electronic structure and electrical properties of surfaces, interfaces, thin films...
% 74.25.Uv 	Vortex phases (includes vortex lattices, vortex liquids, and vortex glasses)
% 74.45.+c 	Proximity effects; Andreev reflection; SN and SNS junctions
% 73.21.Fg 	Electron states and collective excitations in Quantum wells

\begin{abstract}

Certain insulating materials with strong spin-orbit interaction can conduct currents along their edges or surfaces owing to the non-trivial topological properties of their electronic band-structure. This phenomenon is somewhat similar to the integer quantum Hall effect of electrons in strong magnetic fields. Topological insulators analogous to the fractional quantum Hall effect are also possible, but have not yet been observed in any material. Here we show that a quantum well made from a topological band insulator such as Bi$_2$Se$_3$ or Bi$_2$Te$_3$, placed in contact with a superconductor, can be used to realize a two-dimensional topological state with macroscopic many-body quantum entanglement whose excitations carry fractional amounts of electron's charge and spin. This fractional topological insulator is a ``pseudogap'' state of induced spinful $p$-wave Cooper pairs, a new strongly correlated quantum phase with possible applications to spintronic devices and quantum computing.

\end{abstract}

\maketitle

The recently discovered two-dimensional topological insulators (TI) with time-reversal (TR) symmetry\cite{Kane2005, Bernevig2006, Konig2007, Zhang2010} are band-insulators related to integer quantum Hall states in which electron spin plays the role of charge. They can be obtained in HgTe, Bi$_2$Te$_3$ and Bi$_2$Se$_3$ quantum wells owing to the strong spin-orbit coupling, and exhibit topologically protected gapless edge states despite the spin non-conservation\cite{Kane2005a}. The properties of quantum wells are linked to the topologically protected surface states of the extensively studied bulk materials\cite{Hasan2010, Qi2010a, Moore2010}.

Instabilities caused by interactions among electrons can establish unconventional quantum states in TIs, with broken symmetries\cite{Fu2008, Seradjeh2009} or topological order\cite{Pesin2010, Rachel2010, Krempa2010, Young2008}. These envisioned forms of quantum matter could realize robust macroscopic entanglement between spatially separated electrons in the TI materials, which motivates both the fundamental research and the quest for applications in spintronics and quantum computing. Here we aim to realize a new class of strongly correlated TIs that exhibit phenomena reminiscent of the fractional quantum Hall effect (FQHE) in strong magnetic fields, but without its TR symmetry violation\cite{Levin2009, Karch2010, Cho2010, Maciejko2010, Swingle2011, Park2011, Santos2011, Neupert2011}. Such fractional TIs feature quasiparticles that carry fractional amounts of electron's charge and spin. Exotic states with non-Abelian statistics are also possible and promise the ability to perform quantum computation with a greater level of quantum control than in FQHE qubits, because both charge and spin can be manipulated and entangled.

One approach to obtaining fractional TIs, inspired by the FQHE, exploits Coulomb interactions among electrons in a partially populated band made narrow by the spin-orbit coupling\cite{Sun2011, Neupert2011a, Ghaemi2011}. It might be very difficult to find TI materials with sufficiently narrow bands and strong interactions, so the goal of this paper is to propose a different approach. Here we consider a heterostructure device in which a two-dimensional electron gas can be tuned near a quantum critical point (QCP). Every quantum critical system is sensitive to relevant perturbations that impose their energy scales on dynamics and define the phases that surround the critical point in the phase diagram. We will show that the spin-orbit coupling is characterized by a large ``cyclotron'' energy, and thus indeed represents a relevant perturbation that can dominate near the critical point and stabilize fractional topological states just like a strong magnetic field would. The proposed heterostructure is not only routinely achievable, but also provides the best platform to experimentally seek a variety of topologically non-trivial superconducting and insulating quantum states that have not been observed or hypothesized before, and whose existence is guaranteed by the fundamental principles discussed here.

\begin{figure}
\centering
\includegraphics[height=1.5in]{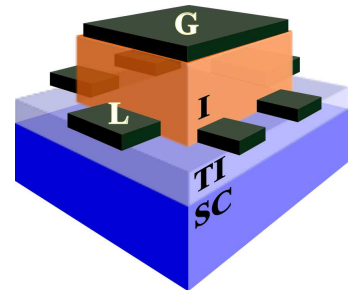}
\caption{\label{Device}The heterostructure device that can host fractional TR-invariant quantum states. A topological insulator (TI) quantum well is sandwiched between a conventional superconductor (SC) and a conventional insulator (I). The gate (G) voltage can be used to control the state of the TI, and the topological properties of the TI can be probed via a Hall-bar setup of leads (L).}
\end{figure}

\begin{figure}
\subfiguretopcaptrue\subcapnoonelinetrue\centering
\def\subfigcapmargin{0.4in}
\raisebox{-0.73in}{
\begin{minipage}{1.7in}
\subfigure[(a)]{\includegraphics[width=1.3in]{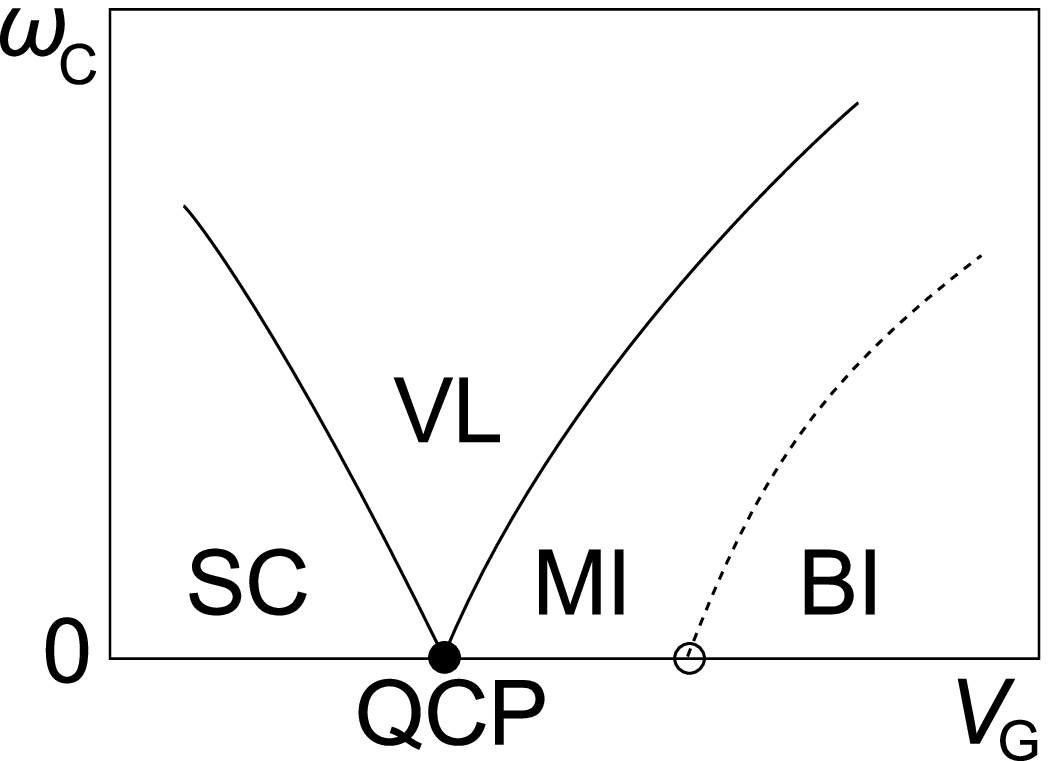}}
\subfigure[(c)]{\includegraphics[width=1.7in]{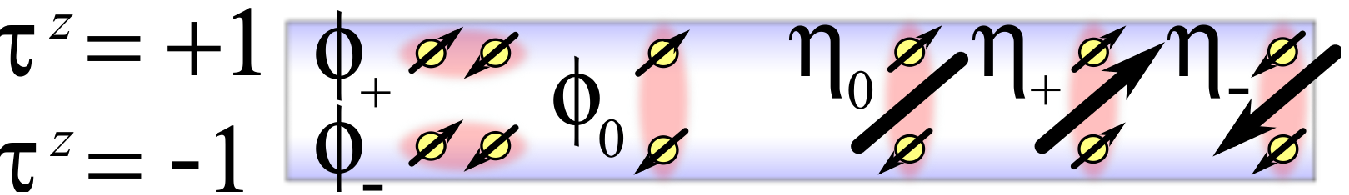}}
\end{minipage}}
\hskip 0.15in
\subfigure[(b)]{\includegraphics[height=1.45in]{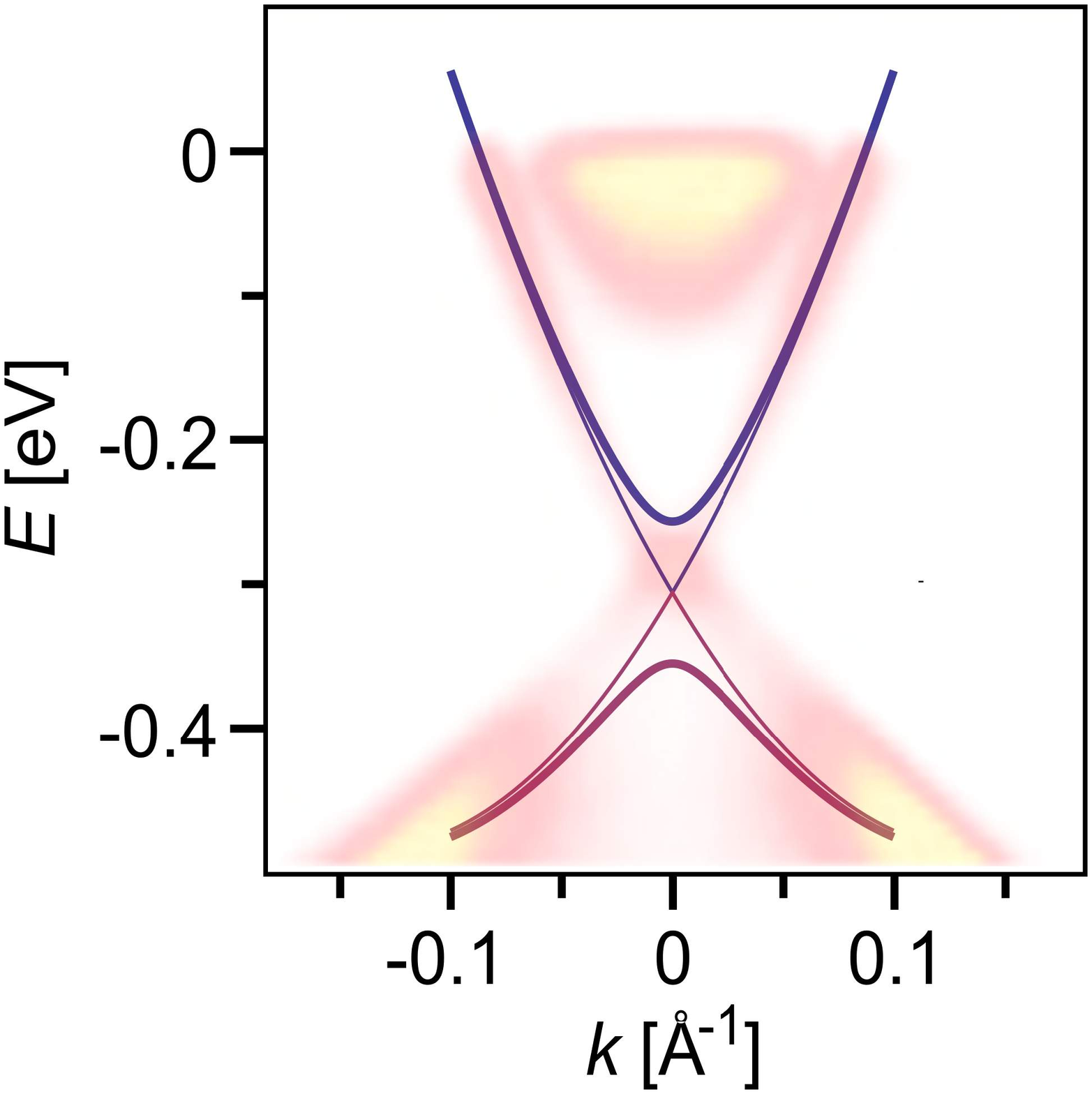}}
\caption{\label{PhDiag}(a) The qualitative zero-temperature phase diagram of attractively interacting electrons in a quantum well. Gate voltage $V_\textrm{G}$ controls the electron gap in the quantum well and tunes the quantum critical point (QCP) between a superconductor (SC) and a bosonic Mott insulator (MI) of Cooper pairs. The lowest energy excitations in the MI are charge $2e$ bosons, but they disappear at the crossover (dashed line) to the band-insulator (BI) where only gapped fermionic quasiparticles with charge $e$ exist. A spin-orbit coupling whose strength is measured by a ``cyclotron'' energy $\omega_{\Phi}$ (defined in the text) introduces a vortex lattice in the superconducting state of spinful $p$-wave Cooper pairs. Quantum melting of such a vortex lattice gives rise to correlated ``vortex liquid'' (VL) states, which are the prime candidates for fractional TIs. (b) The energy spectrum $E(k)$ of the Hamiltonian (\ref{GaugeTh}) for $m\sim 2.5\cdot 10^{-31} \textrm{ kg}$, $v\sim 4\cdot 10^5 \textrm{m/s}$ and $\Delta\sim 100 \textrm{ eV}$ ($k=p/\hbar$). This two-orbital example approximates the ARPES spectrum from the Fig.3d of Ref.\cite{Hsieh2009} when $\Delta=0$. (c) The Cooper pairing channels in the TI include intra-orbital spin singlets ($\Phi_\pm$), inter-orbital spin singlets ($\Phi_0$), and inter-orbital $p$-wave spin triplets ($\eta_m$, where $m\in\lbrace 0,\pm 1\rbrace$ is the $z$-axis spin projection).}
\end{figure}

We engineer a QCP by placing a TI quantum well in contact with a conventional superconductor (SC) as shown in Fig.\ref{Device}. The SC's pairing glue induces a weak short-range attractive interaction between the TI's electrons, but the TI's two-dimensionality assures the formation of bound-state Cooper pairs for any interaction strength \cite{L1977}. Electrons could then be pulled into the TI and immediately bound into pairs by applying a gate voltage, causing a bosonic mean-field quantum phase transition to a superconducting state in the TI \cite{Fisher1989a, Nikolic2010, Nikolic2010b}. The ensuing QCP could naively occur in any pairing channel, but the conventional proximity effect (order parameter leakage) washes out such a QCP of singlets. Only triplet Cooper pairs, made from two electrons in different TI's hybridized surface orbitals, are free to experience a true phase transition if they can be energetically favored. This is where the TI's Rashba spin-orbit coupling steps in. It gives the triplets a crucial boost, and then takes our system away from the QCP as in the Fig.\ref{PhDiag}(a). We will argue that the triplet superconductor in the TI is a vortex lattice of spin-currents, whose quantum melting induced by a gate voltage likely yields a non-Abelian fractional TI.

This scenario can be derived from solid phenomenological arguments alone. We will use the symmetries of the minimal TI model to construct the effective action of the TI affected by the SC. We will then explain why the triplets form a vortex lattice, why such a vortex lattice is inevitably melted by applying a gate voltage, and why the resulting vortex liquid is a candidate for a fractional TI. At the end we discuss the properties of the possible fractional TIs in our system and limitations of our model.

The minimal model Hamiltonian of a non-interacting TI quantum well can be written as \cite{Liu2010a, Shan2010}:
\begin{equation}\label{GaugeTh}
H = \frac{({\bf p} - \tau^z \boldsymbol{\mathcal{A}})^{2}}{2m} + \Delta\tau^{x} - \mu
  \quad,\quad
\boldsymbol{\mathcal{A}} = -mv(\hat{{\bf z}}\times{\bf S}) ~~~ \ .
\end{equation}
It describes four electron states per momentum $\bf p$, labeled by the spin projection $S^z = \pm \frac{1}{2}$ (in the $\hbar=1$ units), and the orbital index $\tau^z = \pm 1$ equivalent to the top/bottom surface of the quantum well. The vector spin operator is ${\bf S} = \frac{1}{2} \sigma^a \hat{\bf r}^a$, $a\in\lbrace x,y,z \rbrace$, and $\sigma^a$ and $\tau^a$ are Pauli matrices that operate on the spin and orbital states respectively. The static Yang-Mills SU(2) gauge field $\boldsymbol{\mathcal{A}}$ embodies the Rashba spin-orbit coupling \cite{Frohlich1992} $H_{\textrm{so}} = v\,\hat{{\bf z}}({\bf S}\times{\bf p})\tau^{z}$ and produces a massless Dirac spectrum if $\Delta=0$. However, inter-surface tunneling $\Delta \neq 0$ opens a bandgap, assuming that the model applies only to momenta $p<\Lambda = \sqrt{(mv)^2-(\Delta/v)^2}$. A natural cut-off $\Lambda$ is provided by the lattice potential in materials. The mass $m$ describes a small Dirac-cone curvature seen in ARPES measurements\cite{Hsieh2009}. Fig.\ref{PhDiag}(b) shows that (\ref{GaugeTh}) adequately approximates materials, with a relatively large fitted $m$. This model has the relativistic particle-hole symmetry when $\mu=0$ and $m\to\infty$. Its many-body ground state is a band-insulator for $|\mu|<|\Delta|$, which is topological when $\Delta$ has a proper sign\cite{Bernevig2006}.

The spin-orbit SU(2) gauge field from (\ref{GaugeTh}) carries a non-zero ``magnetic'' Yang-Mills flux \cite{Peskin1995} ($\mu,\nu... \in \lbrace t,x,y \rbrace$):
\begin{equation}\label{Flux}
\Phi^\mu = \epsilon^{\mu\nu\lambda} ( \partial_\nu \mathcal{A}_\lambda - i \tau^z \mathcal{A}_\nu \mathcal{A}_\lambda )
         = \frac{1}{2}(mv)^{2}\delta_{\mu t}\,\tau^z\sigma^z \ .
\end{equation}
Note that the SU(2) charge $\tau^z$ is required here by gauge invariance. Being a generalization of the U(1) magnetic flux responsible for the Hall effect, the SU(2) flux is the source of topological phenomena in TIs and sets their ``cyclotron'' energy scale $\omega_\Phi=mv^2$. Our construction of the effective action for interacting electrons will greatly benefit from exposing the SU(2) gauge symmetry of the idealized model (\ref{GaugeTh}). At the end, we will discuss the consequences of gauge symmetry violations in real materials.

The electron dynamics in the TI quantum well is altered by the proximity to the SC in the device from Fig.\ref{Device}. The SC is a fully gapped quantum liquid of Cooper pairs characterized by two energy scales, the pairing $\Delta_{\textrm{p}}$ and photon $\Delta_\gamma = \hbar c\lambda_{\textrm{L}}^{-1}$ gaps, where $c$ is the speed of light and $\lambda_{\textrm{L}}$ is the London penetration depth. Fermionic quasiparticles have anomalously small or vanishing density of states below the pairing gap, which is $\Delta_{\textrm{p}} = 2\hbar v_{f}/\pi\xi$ in conventional superconductors with Fermi velocity $v_f$ and coherence length $\xi$. The smaller of the two gaps defines a cut-off energy for the low-energy dynamics in the TI that we shall discuss. The dynamics responsible for the triplet superconductor-insulator transition in the TI is indeed defined below this cut-off and hence can be captured by a two-dimensional effective theory whose degrees of freedom are decoupled from those of the SC. We will show that the resulting theory indeed features a triplet superconductor-insulator transition inside the TI across which $\Delta_{\textrm{p}} \neq 0$.

\begin{table}
\centering
  \begin{tabular}{|c||c|c|c|}
    \hline
    & $\psi_{\tau\sigma}({\bf k})$ & $\phi_{n}({\bf k})$ & $\eta_{m}({\bf k})$ \\
    \hline\hline
    $\mathcal{T}_{{\bf r}}$ translations & $\psi_{\tau\sigma}({\bf k})$
    & $\phi_{n}({\bf k})$ & $\eta_{m}({\bf k})$ \\ \hline
    $\mathcal{R}_{\theta}$ rotations & $\psi_{\tau\sigma}(\mathcal{R}_{\theta}{\bf k})$
    & $\phi_{n}(\mathcal{R}_{\theta}{\bf k})$ & $\eta_{m}(\mathcal{R}_{\theta}{\bf k})$ \\ \hline
    $\mathcal{R}_{i}$ spatial reflect. & $\psi_{\tau\bar{\sigma}}(\mathcal{R}_{i}{\bf k})$
    & $-\phi_{n}(\mathcal{R}_{i}{\bf k})$ & $\eta_{\bar{m}}(\mathcal{R}_{i}{\bf k})$ \\ \hline
    $\mathcal{I}_{t}$ time reversal & $\sigma\psi_{\tau\bar{\sigma}}^{\dagger}(-{\bf k})$
    & $-\phi_{n}^{\dagger}(-{\bf k})$ & $(-1)^m\eta_{\bar{m}}^{\dagger}(-{\bf k})$ \\ \hline
    $\mathcal{C}$ charge U(1) & $e^{i\theta}\psi_{\tau\sigma}({\bf k})$
    & $e^{i2\theta}\phi_{n}({\bf k})$ & $e^{i2\theta}\eta_{m}({\bf k})$ \\ \hline
    $\mathcal{S}$ spin U(1) & $e^{i\sigma\theta}\psi_{\tau\sigma}({\bf k})$
    & $\phi_{n}({\bf k})$ & $e^{i2m\theta}\eta_{m}({\bf k})$ \\ \hline
    local spin SU(2) & $W_{\sigma\sigma'}\psi_{\tau\sigma'}({\bf k})$
    & $\phi_{n}({\bf k})$ & $U_{mm'}\eta_{m'}({\bf k})$ \\ \hline
  \end{tabular}
\caption{\label{SymTransf}The symmetry transformations of electron $\psi_{\sigma\tau}$, singlet $\phi_n$ and triplet $\eta_m$ fields in (\ref{HS}). $W$ and $U$ are SU(2) transformation matrices $\exp\left(i\gamma^a\theta^a\right)$ with SU(2) generators $\gamma^a$, $a\in\lbrace x,y,z \rbrace$ in the $S=\frac{1}{2}$ and $S=1$ representations respectively. ($\gamma^a$ are related to the spin matrices $S^a = \hbar \gamma^a$. Also, $\tau,\sigma = \pm 1$; $n,m \in \lbrace \pm 1,0 \rbrace$; $\bar{l} \equiv -l$.)}
\end{table}

Our effective TI model is given by the imaginary-time action $S = \int \dd\widetilde{t} \, \dd r^2 \, \psi^\dagger (\partial_0 + H) \psi + S_{\textrm{int}}$. Living near the conventional SC, all TI's electrons couple to its phonons and thus acquire BCS-like short-range attractive interactions among themselves, irrespective of their spin or orbital state. This is generic, but overcoming the Coulomb repulsion in the TI requires a sufficiently strong pairing in the SC and a sufficiently thin quantum well. Without knowing the microscopic form and strength of these interactions, we must consider all channels:
\begin{eqnarray}\label{Interactions}
&& \!\!\!\!\!\!\!\! S_{\textrm{int}} = \frac{1}{2} \int \dd\widetilde{t} \, \dd^2 r \, \Bigl(
    U_1 \psi_{\tau\sigma}^\dagger\psi_{\tau\sigma'}^\dagger
        \psi_{\tau\sigma'}^{\phantom{\dagger}}\psi_{\tau\sigma}^{\phantom{\dagger}} \\ && ~~~~
   +U_2 \psi_{\tau\sigma}^\dagger\psi_{\bar{\tau}\sigma'}^\dagger
        \psi_{\bar{\tau}\sigma'}^{\phantom{\dagger}}\psi_{\tau\sigma}^{\phantom{\dagger}}
   +U_3 \psi_{\tau\sigma}^\dagger\psi_{\bar{\tau}\sigma'}^\dagger
        \psi_{\tau\sigma'}^{\phantom{\dagger}}\psi_{\bar{\tau}\sigma}^{\phantom{\dagger}} \Bigr) + \cdots \nonumber
\end{eqnarray}
Here $\tau = \pm 1$ and $\sigma = \pm 1$ label the orbital $\tau^z$ and spin $S^z$ states of the electron fields $\psi_{\tau\sigma}$  respectively ($\bar{\tau}=-\tau$), while the dots denote weak orbital-non-conserving forces. By applying the Hubbard-Stratonovich transformation on the path-integral, we can eliminate the interaction couplings (\ref{Interactions}) in favor of six Cooper pair fields displayed in Fig.\ref{PhDiag}(c): two intra-orbital singlets $\phi_\pm$ ($U_1$), two inter-orbital $S^z=\pm 1$ triplets $\eta_\pm$ ($U_{2/3}$ at $\sigma\!\!=\!\!\sigma'$), inter-orbital singlet $\phi_0$ and $S^z=0$ triplet $\eta_0$ ($U_{2/3}$ at $\sigma\!\!\neq\!\!\sigma'$):
\begin{eqnarray}\label{HS}
&& \!\!\!\!\!\! S_{\textrm{int}}' = \int \dd\widetilde{t} \, d^{2}r \, \Biggl\lbrace
  \sum_{\tau=\pm 1} \left( u|\phi_{\tau}^{\phantom{\dagger}}|^{2}
    + \phi_{\tau}^{\phantom{\dagger}} \epsilon_{\sigma\sigma'}^{\phantom{\dagger}}
      \psi_{\tau\sigma}^{\dagger}\psi_{\tau\sigma'}^{\dagger} + h.c.\right) \nonumber \\
&& ~~ + u'|\phi_0^{\phantom{\dagger}}|^{2} + \phi_0^{\phantom{\dagger}} \frac{1}{\sqrt{2}}
           \left( \psi_{+\uparrow}^{\dagger}\psi_{-\downarrow}^{\dagger}
          -\psi_{+\downarrow}^{\dagger}\psi_{-\uparrow}^{\dagger} \right) + h.c. \nonumber \\
&& ~~ + \sum_{\sigma=\pm 1} \left( v |\eta_\sigma^{\phantom{\dagger}}|^{2} + \eta_\sigma^{\phantom{\dagger}}
             \psi_{+\sigma}^{\dagger}\psi_{-\sigma}^{\dagger} + h.c. \right) \\
&& ~~ + v'|\eta_0^{\phantom{\dagger}}|^{2} + \eta_0^{\phantom{\dagger}} \frac{1}{\sqrt{2}}
           \left( \psi_{+\uparrow}^{\dagger}\psi_{-\downarrow}^{\dagger}
          +\psi_{+\downarrow}^{\dagger}\psi_{-\uparrow}^{\dagger} \right) + h.c. \Biggr\rbrace \nonumber
\end{eqnarray}
The symmetry transformations of these fields are summarized in the Table~\ref{SymTransf}. In conjunction with (\ref{GaugeTh}), the SU(2) symmetry would imply $v=v'$.

Fermionic excitations remain gapped across superconductor-insulator quantum phase transition in simple two-dimensional band-insulators with attractive interactions\cite{Nikolic2010}, and we will explain shortly why this also holds in TIs. Then, we may integrate-out the gapped fermion fields in the path-integral to obtain a purely bosonic effective action $S_{\textrm{eff}}$ that describes Cooper pair dynamics at energies below the pairing gap. We can avoid a complicated calculation by relying on symmetries to construct the Landau-Ginzburg form of $S_{\textrm{eff}}$. Since two electrons with the same spin from different orbitals have the same cyclotron chirality, the Cooper pairs $\eta_\pm$ with $S^z=\pm 1$ possess the SU(2) charge, unlike the singlet fields. Together with $\eta_0$ they form a triplet $\eta = (\eta_-, \eta_0, \eta_+)$ that minimally couples to the same SU(2) gauge field $\boldsymbol{\mathcal{A}}$ as (\ref{GaugeTh}) but expressed in the $S=1$ representation. This can be seen from the local SU(2) transformations in the Table \ref{SymTransf}. Therefore,
\begin{eqnarray}\label{LG}
&& \!\!\!\!\!\!\! S_{\textrm{eff}} = \int \dd\widetilde{t} \; \dd^2 r \; \Bigl\lbrace
   \phi^\dagger\partial_0\phi
        +\left(\boldsymbol{\nabla}\phi\right)^{\dagger}\hat{K}_{\textrm{s}}\left(\boldsymbol{\nabla}\phi\right)
        +\phi^{\dagger}\hat{t}_{\textrm{s}}\phi \nonumber \\
&& \!\!\!\! +\eta^\dagger\partial_0\eta
             +K_{\textrm{t}}\Bigl\vert\left(\boldsymbol{\nabla}-i\boldsymbol{\mathcal{A}}\right)\eta\Bigr\vert^2
        +\Bigl(t_{\textrm{t}}+\phi^{\dagger}\hat{t}_{\textrm{s}}'\phi\Bigr)|\eta|^2
        +U_t|\eta|^4 \nonumber\\
&& \!\!\!\! +U_{\textrm{s},\sigma_{1}\sigma_{2}\sigma_{3}\sigma_{4}}^{\phantom{\dagger}}
         \phi_{\sigma_{1}}^{\dagger} \phi_{\sigma_{2}}^{\dagger}
         \phi_{\sigma_{3}}^{\phantom{\dagger}} \phi_{\sigma_{4}}^{\phantom{\dagger}}
        -\Delta_{\textrm{s}}^\dagger\phi-\Delta_{\textrm{s}}^{\phantom{\dagger}}\phi^\dagger \Bigr\rbrace \ .
\end{eqnarray}
Some Cooper pair modes may have energy in the two-electron continuum, and should be expelled from $S_{\textrm{eff}}$. We omitted Coulomb interactions, and used the most general non-relativistic dynamics. We organized the singlet fields into a vector $\phi = (\phi_-, \phi_0, \phi_+)$ and wrote their quadratic couplings in the matrix form. The vector $\Delta_{\textrm{s}}$ depends on the SC's order parameter and the SC-TI interface. The singlet matrices $\hat{K}_{\textrm{s}}, \hat{t}_{\textrm{s}}, \hat{t}_{\textrm{s}}'$ and tensor $\hat{U}_{\textrm{s}}$ are TR-invariant, and realistic SU(2) symmetry violations can be captured by additional triplet couplings.

Inter-orbital triplets compete with singlets. One of the intra-orbital singlet channels has a stronger induced interaction than all inter-orbital channels for geometric reasons, which naively means that singlets should condense before triplets when electrons are drawn into the TI from the SC by the gate voltage. Here we neglect the intrinsic singlet condensation due to $\Delta_{\textrm{s}}\neq 0$, made small by the TI's bandgap. However, the Rashba spin-orbit coupling in $\mathcal{A}_\mu$ mixes the triplets into two helical modes, analogous to the Dirac conduction and valence band eigenstates of (\ref{GaugeTh}). One helical mode has energy that decreases when its momentum grows (like the Dirac valence band), and thus ``always'' condenses at sufficiently large momenta according to (\ref{LG}). It has a natural advantage over singlets despite its origin in the weaker induced interaction.

\begin{figure}
\centering
\includegraphics[height=1.2in]{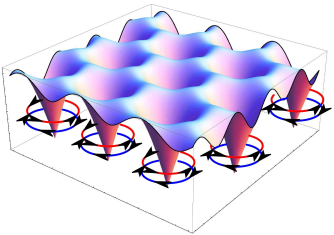}
\caption{\label{SU2vl}A TR-invariant Abrikosov lattice in the helical triplet condensate, $\eta_+^{ } = \eta_-^*$, $\eta_0^{ } = -\eta_0^*$. Coinciding vortices in $\eta_+^{ }$ (red circles) and antivortices in $\eta_-^{ }$ (blue circles), comprise an equilibrium state without charge currents and spin texture, which gains energy by its Rashba-coupled spin current loops.}
\end{figure}

The helical condensate locally gains Rashba energy $\hat{\bf z}({\bf S} \times {\bf p}) < 0$ in (\ref{LG}) by coupling to $\mathcal{A}_\mu$ the TR-invariant currents of properly oriented spin (perpendicular to the current flow). Such a state is globally in equilibrium only if the currents flow in loops. The optimal configuration is always a vortex lattice \cite{Abrikosov1957, Supplement}, illustrated in Fig.\ref{SU2vl}, and its existence also gives birth to fractional TIs. Imagine tuning the gate voltage to reduce the superconducting stiffness $\rho_{\textrm{s}}$ toward zero. The vortex kinetic energy due to zero-point quantum motion can be estimated from the Heisenberg uncertainty as $E_{\textrm{kin}} \sim l_{\Phi}^{-2} / m_{\vv}^{\phantom{2}}$, where $l_{\Phi}$ is the SU(2) ''magnetic length'', and $m_{\vv}$ is the effective vortex mass. In (charged) superconductors, $m_{\vv}$ is roughly constant as $\rho_{\textrm{s}}\to 0$, but turns into $m_{\vv} \sim |\log(\rho_{\textrm{s}})|$ when the screening length $\lambda_{\textrm{L}} \sim \rho_{\textrm{s}}^{-1/2}$ diverges\cite{nikolic:134511}. The potential energy due to the vortex lattice stiffness scales as $E_{\textrm{pot}} \sim \rho_{\textrm{s}}$ per vortex ($\rho_{\textrm{s}}^2$ if the spectrum has Landau levels), which easily follows from the free energy expansion\cite{nikolic:144507} in powers of $\rho_{\textrm{s}}$. There is a critical \emph{finite} $\rho_{\textrm{s}}$ at which the vortex lattice melts in a first order transition because $E_{\textrm{kin}} \ge E_{\textrm{pot}}$. This happens at the solid line that separates the SC and VL regions in Fig.\ref{PhDiag}(a). Since $\rho_{\textrm{s}}$ also measures the quasiparticle pairing gap, its finite value implies that the transition is shaped by the Cooper pair dynamics below the fermion excitation gap. The resulting insulator is a quantum vortex liquid of uncondensed Cooper pairs, whose qualitative properties are captured by the purely bosonic theory (\ref{LG}).

Quantum liquids of SU(2) vortices are the prime candidates for fractional TIs when their density is comparable or larger than the density of Cooper pairs (otherwise, Mott or density-wave insulators are stable). This expectation is based on the transitions from vortex lattice condensates to fractional quantum Hall states in the analogous system of bosons in (effective) magnetic fields\cite{Wilkin2000, Cooper2001, Regnault2003, Chang2005, Cooper2008}. The mass $m$ in (\ref{GaugeTh}) can be estimated from the curvatures of the Dirac cones in ARPES experiments\cite{Hsieh2009}, and it is larger than the ``spin-orbit'' mass $m_{\textrm{so}} = \Delta/v^{2}$ by a factor of $\lambda = \frac{m}{m_{\textrm{so}}} \approx 5-10$ ($v \approx 5 \cdot 10^5 \textrm{ m/s}$). The cyclotron energy $\omega_\Phi = \Phi/m = \lambda \Delta$ is not small in quantum wells with bandgaps $\Delta = 10-100 \textrm{ meV}$ that can be engineered with a few quintuple layers\cite{Zhang2010}. The density of ``magnetic'' SU(2) flux quanta is $n_\Phi = \Phi/h^2 = \lambda^2 \Delta^2/(vh)^2 \approx \lambda^2 \times 2 \cdot 10^{15} \textrm{ m}^{-2}$. These estimates look promising if we compare them with typical flux-quantum densities $n_{\phi} = B(hc/e)^{-1} \approx 2.5 \cdot 10^{15} \textrm{ m}^{-2}$ (in $B=10 \textrm{ T}$) and cyclotron scales $\omega_\phi = \hbar eB/mc \approx 1 \textrm{ meV}$ of electrons in fractional quantum Hall states. The TI's Cooper pair density is controlled by the gate voltage, and can be brought near and below $n_\Phi$ to stabilize a \emph{fractional incompressible quantum liquid} in a finite parameter range of size $\omega_\Phi$ surrounding the QCP in Fig.\ref{PhDiag}(a). Detecting fractional charge and statistics in the absence of magnetic fields, by shot-noise or quantum interferometry methods from FQHE experiments\cite{de-Picciotto1997, Camino2007}, would provide clear evidence of an established fractional TI in the quantum well.

Without microscopic modeling and experimental data we cannot rule out a possibility that singlets would condense before triplets in a particular device. But even then, a further raise of the gate voltage would eventually condense triplets. Singlets cannot completely screen out the gate from triplets because they repel each other stronger than they repel the triplets, by the Pauli exclusion principle. Future experimental probes of topological spin dynamics may be able to reveal fractional $\eta$ vortex liquids even if they coexisted with a singlet superconducting state of the $\phi$ fields (which cannot screen spin).

Finding the precise nature of the fractional TIs goes beyond the scope of this paper as it requires the exact diagonalization of a microscopic model. Instead, we can illustrate their bosonic character by a simple example, such as the bosonic Laughlin wavefunction \cite{Santos2011} of $2N$ triplet Cooper pairs $\eta_\pm$ whose coordinates are $z_{i\pm}$:
\begin{equation}
\Psi = \prod_{i<j}^{1\dots N} \left( z_{i+}^{\phantom{*}} - z_{j+}^{\phantom{*}} \right)^n
  \left( z_{i-}^* - z_{j-}^* \right)^n
  \prod_{i}^{1\dots N} e^{-\frac{|z_{i+}|^2 + |z_{i-}|^2}{4l^2}} \ . \nonumber
\end{equation}
The integer $n$ is even, and this Abelian TR-invariant state has excitations with fractional charge $2e/n$, spin $\hbar/n$ and spin-Hall conductivity $\sigma_{xy}^{\textrm{s}} = 4e\hbar/(nh)$. Since $\langle |\eta_\pm|^2 \rangle \equiv \Phi/(2\pi n)$ and $\Phi$ can be calculated from (\ref{LG}) and (\ref{Flux}), one can find $n$ in any ground state and identify Laughlin states by integer-valued $n$. The wavefunctions of hierarchical quantum spin-Hall states can also be constructed \cite{Santos2011, Neupert2011}. They all describe TR-invariant vortex liquids of spinful bosons (with vortex density $l^{-2}$), and thus are not far from being good candidates for the fractional TIs in our system. However, they are not adequate either because the $S^z$ spin component is not conserved. It is presently unknown how to write a proper wavefunction for a fractional TI shaped by the Rashba spin-orbit coupling, but an effective field-theory description is available and points to the naturally non-Abelian character of the ensuing incompressible quantum liquids. \cite{Supplement, Nikolic2011, Nikolic2012}

Instead of $S^z$, the spin quantum number in an ideal Rashba-based TI is the eigenvalue of $\hat{\bf{z}}(\bf{S}\times\hat{\bf{p}})$ as evident from (\ref{GaugeTh}). If it were conserved, measuring its average on the fractional TI's quasiparticles in the momentum $\bf{p}$ eigenstate would yield a fraction of $\pm\hbar$. However, the realistic complete spin non-conservation, manifested as a gauge symmetry violation in (\ref{LG}), spoils the measurements of fractional spin. At least there is no obstacle to observing the conserved fractional charge, so the fractional TIs can exist. The fractional spin is a degree of freedom rather than a quantum number of quasiparticles (which has a mixed spin and orbital character). Combining an integer number of fractional quasiparticles must reconstitute a triplet Cooper pair, so the quasiparticles must inherit from it a degree of freedom that transforms like spin under time-reversal and spans multiple basis states. Its fractional quantization is guaranteed by the fundamental properties of vortex dynamics in incompressible quantum liquids, and its spin-orbit coupling may yield new topological orders not found in FQHE systems.

We are very grateful to Michael Levin for insightful discussions, and to the Aspen Center for Physics for its hospitality. This research was supported by the Office of Naval Research (grant N00014-09-1-1025A), the National Institute of Standards and Technology (grant 70NANB7H6138, Am 001), and the U.S. Department of Energy, Office of Basic Energy Sciences, Division of Materials Sciences and Engineering under Award DE-FG02-08ER46544.

% We are very grateful to Michael Levin for insightful discussions, and to the Aspen Center for Physics for its hospitality. This research was supported by the Office of Naval Research, the National Institute of Standards and Technology, and the U.S. Department of Energy.

%\bibliographystyle{/usr/share/texmf/bibtex/bst/revtex/prsty}
%\bibliography{/home/dasko/Science/Bibliography/references}

%\begin{comment}

%\end{comment}

\newpage

\def \LG {5}
\def \TFT {46,47}
\def \VL {3}

\begin{center}
  \Large Supplemental material
\end{center}

This supplement provides technical support for our central phenomenological claim: the condensate of spinful triplet Cooper pairs in the presence of the Rashba spin-orbit coupling is a vortex lattice. This claim is fully based on two simple facts that we outlined in the main article: 1) any external gauge flux ``diamagnetically'' encourages current flow, and 2) any current flow in equilibrium superconductors must take the form of a vortex lattice. The main purpose of the following calculations is to illustrate these facts. A by-product of these calculations, not essential for our main analysis, is the likely form of the triplet vortex lattice involving ``helical'' spin-currents (whose spin is perpendicular to its flow by the right-hand rule). At the end, we will briefly discuss the nature of incompressible quantum liquids obtained by the quantum melting of such a triplet vortex lattice.

We performed the calculations below in the continuum limit to keep them tractable and to analytically reveal the possible character of vortices. However, it should be kept in mind that our system of interest lives on a lattice and has a momentum cut-off $\Lambda'$ smaller than or comparable to the characteristic spin-orbit momentum $\Lambda = \sqrt{(mv)^2-(\Delta/v)^2}$ at which the naive continuum-limit dispersion reaches minimum. The dynamics at the cut-off scales is important in this problem, and will be adequately studied in the future.

\section{Helical triplet condensate}\label{secHC}

A type-II superconductor generally forms a vortex lattice in the presence of a uniform flux density that cannot be expelled from the system. This fact is well-known in the case of U(1) magnetic fields, and here we show by a simple calculation that it also holds for the Rashba spin-orbit SU(2) fluxes. We will use the conservation laws for spin-currents to restrict the ground state order parameter of a triplet superconductor with Rashba spin-orbit coupling. This will allow us to construct the qualitative pattern of spin-currents in the ensuing vortex lattice.

Consider the action of triplet Cooper pairs in the topological insulator (TI) quantum well, extracted from the {Eq.\LG} in the main article:
\begin{eqnarray}\label{Act}
S_{\textrm{eff}} \!\!&=&\!\! \int \dd\widetilde{t} \; \dd^2 r \; \Bigl\lbrace
  \eta^\dagger\partial_0\eta
     +K_{\textrm{t}}\Bigl\vert\left(\boldsymbol{\nabla}-i\boldsymbol{\mathcal{A}}\right)\eta\Bigr\vert^2 \nonumber \\
&&   +t_{\textrm{t}}'|\eta|^2+U_{\textrm{t}}|\eta|^4+\cdots \Bigr\rbrace \ .
\end{eqnarray}
The SU(2) gauge field $\mathcal{A}_\mu = (\mathcal{A}_t , \boldsymbol{\mathcal{A}})$:
\begin{equation}\label{Rashba}
\boldsymbol{\mathcal{A}} = -mv(\hat{{\bf z}}\times{\bf S}) \quad,\quad \mathcal{A}_t = 0 \quad ,
\end{equation}
where $\bf S$ is the spin operator of triplet Cooper pairs (in the $S=1$ representation), embodies the Rashba spin-orbit coupling because the gauged gradient term in the action expands into:
\begin{eqnarray}\label{Kin}
&& \!\!\!\!\!\!\! S_{\textrm{eff}}^{(\textrm{kin})} = \int \dd\widetilde{t} \; \dd^2 r \; \Bigl\lbrace
  K_{\textrm{t}} |\boldsymbol{\nabla} \eta|^2
    + 2K_{\textrm{t}} \eta^\dagger \boldsymbol{\mathcal{A}} (i\boldsymbol{\nabla}) \eta \Bigr\rbrace \\
&& \!\!\!\! = \int \dd\widetilde{t} \; \dd^2 r \; \Bigl\lbrace K_{\textrm{t}} |\boldsymbol{\nabla} \eta|^2
   + 2mv K_{\textrm{t}} \eta^\dagger \hat{{\bf z}} ({\bf S} \times {\bf p}) \eta \Bigr\rbrace \nonumber
\end{eqnarray}
apart from an uninteresting contribution to the mass $t_{\textrm{t}}'$ in (\ref{Act}). We showed in the main article that the gauge field such as (\ref{Rashba}) has a finite SU(2) flux $\Phi \neq 0$:
\begin{equation}\label{Flux}
\Phi^\mu = \epsilon^{\mu\nu\lambda} ( \partial_\nu \mathcal{A}_\lambda - i \mathcal{A}_\nu \mathcal{A}_\lambda )
         = \Phi \delta_{\mu t} S^z \ .
\end{equation}

The charge $j_\mu^{\phantom{a}}$ and spin $J_\mu^a$ current densities of the triplet fields are:
\begin{eqnarray}\label{Currents}
j_{t} \!\!&=&\!\! \eta^{\dagger}\eta \\
j_{i} \!\!&=&\!\! -\frac{i}{2m}\Bigl(\eta^{\dagger}(\partial_{i}\eta)-(\partial_{i}\eta^{\dagger})\eta\Bigr)\nonumber\\
J_{t}^{a} \!\!&=&\!\! \eta^{\dagger}S^{a}\eta\nonumber\\
J_{i}^{a} \!\!&=&\!\! -\frac{i}{2m}\Bigl(\eta^{\dagger}S^{a}(\partial_{i}\eta)-(\partial_{i}\eta^{\dagger})S^{a}\eta\Bigr)
                       \nonumber
\end{eqnarray}
where the temporal current components ($\mu=t$) are ordinary densities of charge and spin, the spatial current components ($\mu\equiv i\in\lbrace x,y \rbrace$) indicate the flow of charge and spin, braces denote anti-commutators, and $S^{a}$, $a\in\lbrace x,y,z\rbrace$ are the spin projection operators:
\begin{eqnarray}
&&
  S^x = \frac{1}{\sqrt{2}}\left(\begin{array}{ccc}0&1&0\\1&0&1\\0&1&0\end{array}\right) \quad,\quad
  S^y = \frac{1}{\sqrt{2}}\left(\begin{array}{ccc}0&-i&0\\i&0&-i\\0&i&0\end{array}\right) \nonumber \\
&& ~~~~~~~~~~~~~~~~~~~
  S^z = \left(\begin{array}{ccc}1&0&0\\0&0&0\\0&0&-1\end{array}\right) \ .
\end{eqnarray}

The most general TR-invariant triplet order parameter can be expressed as a function of three real functions of coordinates, $\theta$, $\alpha$ and $\zeta>0$:
\begin{equation}\label{OP}
\eta = \left(\begin{array}{c} \eta_{+} \\ \eta_{0} \\ \eta_{-} \end{array}\right)
     = \zeta e^{i\theta_0}
       \left(\begin{array}{c} e^{-i\theta}\cos\alpha \\ i\sqrt{2}\sin\alpha\\ e^{i\theta}\cos\alpha\end{array}\right) \ .
\end{equation}
No physical observable depends on the arbitrary constant phase $e^{i\theta_0}$, and we may treat $\theta$ and $\alpha$ as two independent angles in the $(0,2\pi)$ interval even though the transformation $\theta\to\theta+\pi$, $\alpha\to\pi-\alpha$ leaves the $\eta$ spinor unchanged. The non-vanishing current density components (\ref{Currents}) are only those that remain invariant under TR:
\begin{eqnarray}\label{Currents2}
j_t \!\!&=&\!\! 2 \zeta^2 \\[0.1in]
{\bf J}^{x} \!\!&=&\!\! \frac{2\zeta^{2}}{m}\left\lbrack \cos\theta\,\boldsymbol{\nabla}\alpha
  +\frac{1}{2}\sin\theta\sin(2\alpha)\,\boldsymbol{\nabla}\theta\right\rbrack \nonumber \\
{\bf J}^{y} \!\!&=&\!\! \frac{2\zeta^{2}}{m}\left\lbrack \sin\theta\,\boldsymbol{\nabla}\alpha
  -\frac{1}{2}\cos\theta\sin(2\alpha)\,\boldsymbol{\nabla}\theta\right\rbrack \nonumber \\
{\bf J}^{z} \!\!&=&\!\! -\frac{2\zeta^{2}}{m}\cos^{2}\alpha\,\boldsymbol{\nabla}\theta \nonumber \ .
\end{eqnarray}
Specifically, there is no spin-texture ($J_t^a = 0$) and no flow of charge (${\bf j}=0$), but charge density $j_t \neq 0$ is free to break translational symmetry by forming a vortex lattice, and so is the spin-current density ${\bf J}^a$.

\subsection{Restrictions on the order parameter}

The superconductor's order parameter is a classical quantity by the virtue of being the expectation value of a field operator. Therefore, the currents obtained from the order parameter must obey the classical conservation laws. These laws can be derived from the equation of motion, which for a non-relativistic theory like ours is the Schrodinger equation (or its adjoint):
\begin{eqnarray}
\frac{1}{2m}(-i\boldsymbol{\nabla}-\boldsymbol{\mathcal{A}})^{2}\eta+\cdots \!\!&=&\!\!
  i\frac{\partial\eta}{\partial t} \\
\frac{1}{2m}\Bigl\lbrack(-i\boldsymbol{\nabla}-\boldsymbol{\mathcal{A}})^{2}\eta\Bigr\rbrack^\dagger+\cdots \!\!&=&\!\!
 -i\frac{\partial\eta^{\dagger}}{\partial t} \ . \nonumber
\end{eqnarray}
The conservation laws for spin-currents are obtained when the second equation is multiplied from right by $S^{a}\eta$ and subtracted from the first equation multiplied from left by $\eta^{\dagger}S^{a}$. After some algebraic manipulation, one arrives at:
\begin{equation}\label{CurCons}
  \partial_\mu I_\mu -i\lbrack\mathcal{A}_\mu,I_\mu\rbrack = 0
\end{equation}
where $I_\mu^{\phantom{a}} = I_\mu^a S^a$ are the SU(2) matrices of the gauge covariant spin-currents
\begin{equation}\label{GCCur}
I_t^a = J_t^a \quad,\quad I_i^a = J_i^a - \frac{1}{2m}\eta^\dagger\lbrace S^a,\mathcal{A}_i\rbrace\eta \ .
\end{equation}
The expression (\ref{CurCons}) is perhaps more familiar in its gauge-covariant form $\lbrack D_\mu, I_\mu \rbrack = 0$, where $D_\mu = \partial_\mu - i\mathcal{A}_\mu$ is the covariant derivative. The matrix formulation of spin currents is convenient due to its simple SU(2) transformation:
\begin{equation}
I_\mu^{\phantom{a}} \to W I_\mu^{\phantom{a}} W^\dagger \ .
\end{equation}
The general SU(2) gauge transformation is specified by a matrix $W = \exp(i S^a \theta^a)$ that depends on three angles $\theta^a({\bf r},t)$. It affects the spinor and gauge fields in the following manner:
\begin{equation}
\eta \to W \eta \quad,\quad
\mathcal{A}_\mu \to W \mathcal{A}_\mu W^\dagger + i W \partial_\mu W^\dagger \ ,
\end{equation}
and leaves the spin-current conservation law (\ref{CurCons}) invariant. The charge-current conservation law takes the usual form $\partial_\mu j_\mu = 0$.

All time derivatives in the current conservation laws must vanish in equilibrium. Then, the Rashba spin-orbit coupling (\ref{Rashba}) turns (\ref{CurCons}) into:
\begin{eqnarray}\label{CurCons2}
\boldsymbol{\nabla}{\bf I}^{x} \!\!&=&\!\!
   -{\bf A}^{y}{\bf I}^{z}+{\bf A}^{z}{\bf I}^{y}=-mv\, I_{x}^{z}\\
\boldsymbol{\nabla}{\bf I}^{y} \!\!&=&\!\!
   -{\bf A}^{z}{\bf I}^{x}+{\bf A}^{x}{\bf I}^{z}=-mv\, I_{y}^{z}\nonumber\\
\boldsymbol{\nabla}{\bf I}^{z} \!\!&=&\!\!
   -{\bf A}^{x}{\bf I}^{y}+{\bf A}^{y}{\bf I}^{x}=mv(I_{x}^{x}+I_{y}^{y})\nonumber \ .
\end{eqnarray}
The scalar gauge field components $A_\mu^a$ are extracted from $\mathcal{A}_\mu^{\phantom{a}} = A_\mu^a S^a$, and we used the relationship $\lbrack S^a, S^b \rbrack = i\epsilon^{abc} S^c$ between the SU(2) generators $S^a$ in any representation. These equations define constraints that the order parameter must satisfy if it is to be static. In order to express (\ref{CurCons2}) in a relatively compact form, let us define the in-plane spin-currents as double vectors:
\begin{equation}
\vec{\bf J}=\vec{x}\,{\bf J}^{x}+\vec{y}\,{\bf J}^{y} \quad,\quad \vec{\bf I}=\vec{x}\,{\bf I}^{x}+\vec{y}\,{\bf I}^{y} \ .
\end{equation}
The unit-vectors $\vec{x},\vec{y}$ are related to the orientation of spin (by coupling to $S^x, S^y$), as opposed to the unit-vectors $\hat{\bf x},\hat{\bf y}$ that are related to the spatial orientation of current flow. Formally, $\vec{x},\vec{y}$ and $\hat{\bf x},\hat{\bf y}$ live in different vector spaces, while $\vec{\bf J},\vec{\bf I}$ live in both respective vector spaces at the same time. It is also useful to define:
\begin{equation}
\vec{\theta}=\vec{x}\cos\theta+\vec{y}\sin\theta \quad,\quad
  \hat{\boldsymbol{\theta}}=\hat{{\bf x}}\cos\theta+\hat{{\bf y}}\sin\theta \ .
\end{equation}
Then:
\begin{eqnarray}
\vec{\bf J} \!\!&=&\!\! \frac{2\zeta^{2}}{m}\left\lbrack \vec{\theta}\,\boldsymbol{\nabla}\alpha
  -\frac{\sin(2\alpha)}{2}(\vec{z}\times\vec{\theta})\boldsymbol{\nabla}\theta\right\rbrack \\
\vec{\bf I} \!\!&=&\!\! \vec{\bf J}
  +2v\zeta^{2}\biggl\lbrack\cos^{2}\!\alpha\,\vec{\theta}(\hat{{\bf z}}\times\hat{\boldsymbol{\theta}})
  +\sin^{2}\!\alpha\,(\vec{x}\,\hat{{\bf y}}-\vec{y}\,\hat{{\bf x}})\biggr\rbrack \nonumber \ .
\end{eqnarray}
The conservation laws (\ref{CurCons2}) can now be resolved in terms of $\theta$, $\alpha$ and $\zeta$. After some straight-forward algebraic manipulations, one finds:
\begin{eqnarray}\label{CurCons3}
\boldsymbol{\nabla}{\bf J}^{z} \!\!&=&\!\! 4v\zeta^{2}\left(\cos^2\!\alpha\,\boldsymbol{\nabla}\alpha
  +\frac{\sin(2\alpha)}{2}\,\frac{\boldsymbol{\nabla}\zeta}{\zeta}\right)\hat{\boldsymbol{\theta}} \nonumber \\
\vec{\theta}\,\boldsymbol{\nabla}\vec{\bf J} \!\!&=&\!\!
    4v\zeta^{2}\left(\cos^{2}\!\alpha\,\hat{\boldsymbol{\theta}}\boldsymbol{\nabla}\theta
   -\frac{\boldsymbol{\nabla}\zeta}{\zeta}(\hat{\bf z}\times\hat{\boldsymbol{\theta}})\right) \nonumber \\
&& +mv^{2}\zeta^{2}\sin(2\alpha) \\[0.05in]
(\vec{z}\times\vec{\theta})\boldsymbol{\nabla}\vec{\bf J} \!\!&=&\!\! 4v\zeta^{2}\left(
    \frac{\sin(2\alpha)}{2}\boldsymbol{\nabla}\alpha
   +\frac{\boldsymbol{\nabla}\zeta}{\zeta}\sin^{2}\!\alpha\right)\hat{\boldsymbol{\theta}} \nonumber \ .
\end{eqnarray}

\subsection{Energy gain from the Rashba spin-orbit coupling}

The gradient energy of the order parameter (\ref{OP}) extracted from (\ref{Kin}) is:
\begin{eqnarray}\label{Kin2}
\!\!\!\!\!\!\!\! E_{\textrm{kin}} \!\!&=&\!\! \int \dd^2 r \; \Bigl\lbrace K_{\textrm{t}} |\boldsymbol{\nabla} \eta|^2
   + 2m^2 v\, K_{\textrm{t}} \left( J_y^x - J_x^y \right) \Bigr\rbrace \\
\!\!&\propto&\!\! \int \dd^2 r \Bigl\lbrace (\boldsymbol{\nabla}\zeta)^{2}+\zeta^{2}(\boldsymbol{\nabla}\alpha)^{2}
      +\zeta^{2}\cos^{2}\alpha\,(\boldsymbol{\nabla}\theta)^{2} \nonumber \\
&&  +\kappa\zeta^{2}\Bigl\lbrack\sin(2\alpha)\boldsymbol{\nabla}\theta
      -2(\hat{{\bf z}}\times\boldsymbol{\nabla}\alpha)\Bigr\rbrack\hat{\boldsymbol{\theta}}\Bigr\rbrace \nonumber \ ,
\end{eqnarray}
where $\kappa = 2mv$. It plays the decisive role in the spatial modulations of the order parameter. The quadratic gradient terms are pure kinetic energy and always positive. However, the Rashba energy is linear in gradients, so it can be negative and stabilize a spatially inhomogeneous state of non-zero currents.

\begin{figure}
\includegraphics[width=2.5in]{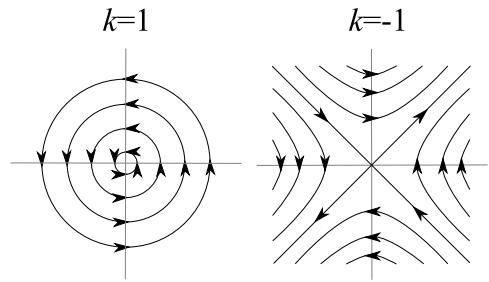}
\caption{\label{VecVortex}Vortices of two-dimensional vector fields ${\bf v} \propto -\hat{\bf x}\sin(k\varphi) + \hat{\bf y}\cos(k\varphi)$, where $\varphi$ is the polar angle, can implement any total rotation angle $2\pi k$ of the vector ${\bf v}$ around a loop that encloses the singularity. However, only $k=1$ can correspond to a current field without any sources at the system boundary. Also, only the $k=1$ case can describe a current field of a quantized U(1) vortex, ${\bf v} = \boldsymbol{\nabla}\theta$. Even if the angle $\theta = n\varphi$ winds by $2\pi n$ on the loop around the singularity, the corresponding vector field ${\bf v} = n(-\hat{\bf x}\sin\varphi + \hat{\bf y}\cos\varphi)$ is still a $k=1$ vector-vortex for any $n$.}
\end{figure}

The first term of the Rashba energy
\begin{equation}
E_{\textrm{R1}} = \kappa \int \dd^2 r \, \zeta^2 \sin(2\alpha)\,\hat{\boldsymbol{\theta}}\boldsymbol{\nabla}\theta
\end{equation}
can provide the maximum energy density gain proportional to $\hat{\boldsymbol{\theta}} \boldsymbol{\nabla}\theta$ for any local value of $\alpha$. Since $\hat{\boldsymbol{\theta}}$ is a unit-vector, the amount of gained energy naively depends on the magnitude of $\boldsymbol{\nabla} \theta$. However, $\boldsymbol{\nabla} \theta \neq 0$ implies rotations of $\hat{\boldsymbol{\theta}}$ along various paths through the system. Assuming that $\alpha$ is approximately homogeneous, a significant $E_{\textrm{R1}}$ energy gain can come only from those regions of space where the vectors $\hat{\boldsymbol{\theta}}$ and $\boldsymbol{\nabla} \theta$ do not rotate with respect to each other. The total angle of $\hat{\boldsymbol{\theta}}$ rotation on a closed loop
\begin{equation}
\oint \dd {\bf l} \boldsymbol{\nabla}\theta = 2\pi n
\end{equation}
must be quantized ($n \in \mathbb{Z}$) in order for the order parameter to be single-valued everywhere in space. In contrast, the vector $\boldsymbol{\nabla} \theta$ always rotates by the total of $0$ or $+2\pi$ on the same loop. The $+2\pi$ angle is obtained whenever $\theta$ has a vortex of \emph{any} non-zero ``charge'' $n$ inside the loop. Therefore, only a single-quantized $n=1$ vortex allows the vectors $\hat{\boldsymbol{\theta}}$ and $\boldsymbol{\nabla} \theta$ to co-rotate and yield a finite spatial average $\langle \hat{\boldsymbol{\theta}} \boldsymbol{\nabla} \theta\rangle$ that can reduce $E_{\textrm{R1}}$. This analysis is illustrated in the Fig.\ref{VecVortex}, and assumes that the loops of interest are simple closed paths without self-intersections.

The second term of the Rashba energy
\begin{equation}\label{ER2}
E_{\textrm{R2}} = -2\kappa \int \dd^2 r \, \zeta^2 (\hat{{\bf z}}\times\boldsymbol{\nabla}\alpha) \hat{\boldsymbol{\theta}}
\end{equation}
behaves better: $\boldsymbol{\nabla}\alpha$ can be large (unlike $\boldsymbol{\nabla}\theta$) while rotating in synchrony with $\hat{\boldsymbol{\theta}}$ along a current flow path. The additional factor of two and the absence of an explicit $\alpha$-dependence also make $E_{\textrm{R2}}$ stronger than $E_{\textrm{R1}}$. We conclude that this part of (\ref{Kin2}) is the main agent in minimizing the ground-state energy. We should, therefore, seek order parameters with $\hat{\boldsymbol{\theta}}$ pointing everywhere in the same direction as $\hat{{\bf z}}\times\boldsymbol{\nabla}\alpha$. This creates a pattern of ``helical'' spin-currents.

The simplest helical current pattern is a uniform flow of the in-plane spin that is perpendicular to the flow direction ($\boldsymbol{\nabla} \theta = 0$, $\boldsymbol{\nabla} \alpha = \textrm{const}$). Such a uniform flow indeed exists as a solution of the stationary Schrodinger equation, which is the equation of motion in our theory. Consequently, this solution satisfies the conservation laws in the bulk. However, it may not qualify as an equilibrium solution at the system boundaries, because it involves spin transport from one side of the system to another. The naive continuum limit considered here circumvents this problem by allowing plane-wave condensates of zero group velocity, which carry no currents and thus do not violate equilibrium. Such condensates must occur at large momenta where the Rashba spin-orbit-coupled triplet Cooper pairs reach their energy minimum. It is, then, extremely plausible that the natural momentum cut-off present in lattice systems could put such states in disadvantage. Our present analysis is not suited for the exploration of dynamics at cut-off scales, but we can nevertheless anticipate the existence of alternative stable lattice condensates.

Our goal here will be to qualitatively reveal the possible nature of alternative condensates using the analytical continuum-limit approach, despite its limitations. An equilibrium pattern of non-zero currents must generally have current loops. Finding them is complicated by the non-trivial conservation laws for spin-currents. We will thus make some approximations to qualitatively deduce the character of loop spin-currents.

\subsection{Coarse-graining approximations}

Since (\ref{ER2}) is the main source of energy gain from having non-vanishing spin-currents, we can focus on its behavior in the well-defined limit of strong Rashba spin-orbit couplings $v$. The characteristic values of $|\boldsymbol{\nabla}\alpha|$ can be large in this limit, leading to rapid spatial variations of $\alpha$. We are then justified in coarse-graining the kinetic energy and currents over the small regions of space where $\alpha$ rotates a full cycle while the other variables change gradually. The kinetic energy (\ref{Kin2}) is approximately:
\begin{eqnarray}\label{Kin3}
E_{\textrm{kin}} \!\!&\propto&\!\! \int \dd^{2}r \biggl\lbrack
  (\boldsymbol{\nabla}\zeta)^{2}+\zeta^{2}(\boldsymbol{\nabla}\alpha)^{2}+\frac{1}{2}\zeta^{2}(\boldsymbol{\nabla}\theta)^{2} \\
&& -2\kappa\zeta^{2}(\hat{{\bf z}}\times\boldsymbol{\nabla}\alpha)\hat{\boldsymbol{\theta}}\biggr\rbrack \ , \nonumber
\end{eqnarray}
while the coarse-grained currents (\ref{Currents2})
\begin{equation}\label{Currents3}
{\bf J}^{z} = -\frac{\zeta^{2}}{m}\,\boldsymbol{\nabla}\theta \quad,\quad
  \vec{\bf J} = \frac{2\zeta^{2}}{m}\,\vec{\theta}\,\boldsymbol{\nabla}\alpha
\end{equation}
obey much simpler coarse-grained conservation laws (\ref{CurCons3}):
\begin{eqnarray}\label{CurCons4}
\boldsymbol{\nabla}{\bf J}^{z} \!\!&=&\!\! 2v\zeta^{2}\,\hat{\boldsymbol{\theta}}\boldsymbol{\nabla}\alpha \\
\vec{\theta}\,\boldsymbol{\nabla}\vec{\bf J} \!\!&=&\!\! 4v\zeta^{2}\left(
   \frac{1}{2}\hat{\boldsymbol{\theta}}\boldsymbol{\nabla}\theta
  -\frac{\boldsymbol{\nabla}\zeta}{\zeta}(\hat{{\bf z}}\times\hat{\boldsymbol{\theta}})\right) \nonumber \\
(\vec{z}\times\vec{\theta})\boldsymbol{\nabla}\vec{\bf J} \!\!&=&\!\! 2v\zeta\,\hat{\boldsymbol{\theta}}\boldsymbol{\nabla}\zeta
  \nonumber \ .
\end{eqnarray}
We can immediately see that $\boldsymbol{\nabla}{\bf J}^z \to 0$ on fairly short length-scales if we keep $\hat{\boldsymbol{\theta}}$ and $\hat{{\bf z}}\times\boldsymbol{\nabla}\alpha$ parallel to each-other to minimize the Rashba energy. Therefore, ${\bf J}^z$ should have no sources or drains. Only current loops in the form of quantized vortices can make ${\bf J}^z$ finite.

Substituting the divergences of (\ref{Currents3}) into (\ref{CurCons4}) yields:
\begin{eqnarray}\label{CurCons5}
-\frac{2\zeta}{m}(\boldsymbol{\nabla}\zeta)(\boldsymbol{\nabla}\theta)-\frac{\zeta^{2}}{m}\nabla^{2}\theta \!\!&=&\!\!
  2v\zeta^{2}\,\hat{\boldsymbol{\theta}}\boldsymbol{\nabla}\alpha\to 0 \nonumber \\
\frac{4\zeta}{m}(\boldsymbol{\nabla}\zeta)(\boldsymbol{\nabla}\alpha)+\frac{2\zeta^{2}}{m}\nabla^{2}\alpha \!\!&=&\!\!
  2v\zeta^{2}\left( \hat{\boldsymbol{\theta}}\boldsymbol{\nabla}\theta
  -\frac{2}{\zeta}(\hat{{\bf z}}\times\hat{\boldsymbol{\theta}})\boldsymbol{\nabla}\zeta\right) \nonumber \\
\frac{2\zeta^{2}}{m}(\boldsymbol{\nabla}\theta)(\boldsymbol{\nabla}\alpha) \!\!&=&\!\!
  2v\zeta\,\hat{\boldsymbol{\theta}}\boldsymbol{\nabla}\zeta
\end{eqnarray}
in the ground state ($\hat{\boldsymbol{\theta}}\boldsymbol{\nabla}\alpha \to 0$). These consequences of spin-current conservation look particularly simple in the regions far away from any vortex singularities, where the spatial variations of the order parameter magnitude $\zeta$ can be neglected:
\begin{eqnarray}\label{CurCons6}
\nabla^{2}\theta \!\!&\to&\!\! 0 \\
\nabla^{2}\alpha \!\!&=&\!\! mv\,\hat{\boldsymbol{\theta}}\boldsymbol{\nabla}\theta \nonumber \\
(\boldsymbol{\nabla}\theta)(\boldsymbol{\nabla}\alpha) \!\!&=&\!\! 0 \nonumber \ .
\end{eqnarray}
Since these conservation laws characterize the ground state, we may use them to learn a great deal about the nature of the order parameter in the helical condensate.

\subsection{The nature of the helical triplet vortex lattice}

We concluded earlier that the currents $\,{\bf J}^z$ should have a certain vortex configuration in the equilibrium minimum-energy state. So, let us understand better the nature of vortices by looking at regions far away from vortex cores. Suppose that $\theta$ winds by $2\pi n$ about the origin, consistent with $\nabla^{2}\theta\to0$ of (\ref{CurCons6}). This makes $\boldsymbol{\nabla} \theta$ tangential to the closed paths around the origin. The last equation of (\ref{CurCons6}) then implies that $\boldsymbol{\nabla} \alpha$ should be aligned radially, in order to be orthogonal to $\boldsymbol{\nabla}\theta$. Such a configuration of the vector field $\boldsymbol{\nabla}\alpha$ must have a symmetric continuous distribution of sources, which could be provided by $\hat{\boldsymbol{\theta}}\boldsymbol{\nabla}\theta\neq0$ in the middle equation of (\ref{CurCons6}). This is clearly a consequence of the finite Rashba spin-orbit coupling ($v\neq 0$), which does not conserve spin in the usual sense. We conclude that $\hat{\boldsymbol{\theta}}$ should keep being parallel to $\boldsymbol{\nabla}\theta$. As we explained earlier, this is possible only if the vortex ``charge'' is $n = +1$.

\begin{figure}
\subfigure[]{\includegraphics[height=1.5in]{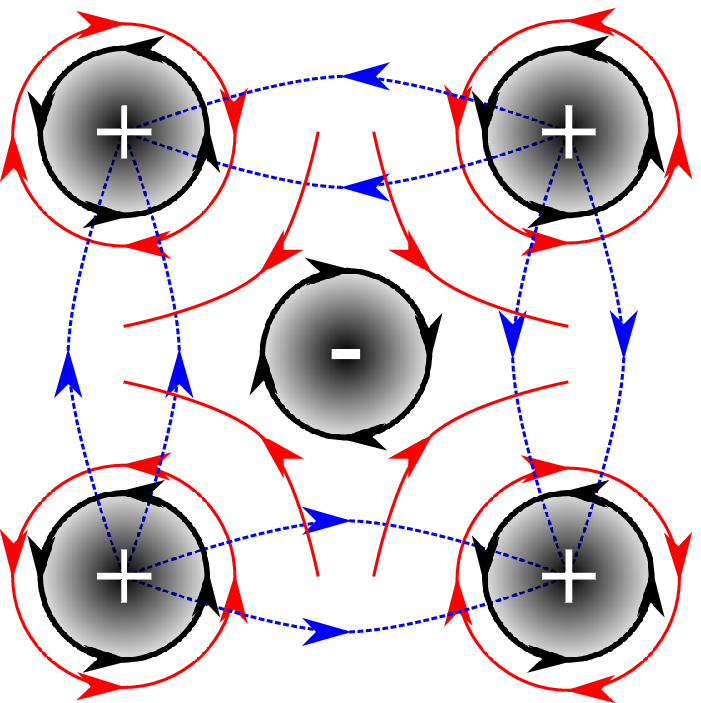}}
\hspace{0.2in}
\subfigure[]{\includegraphics[height=1.5in]{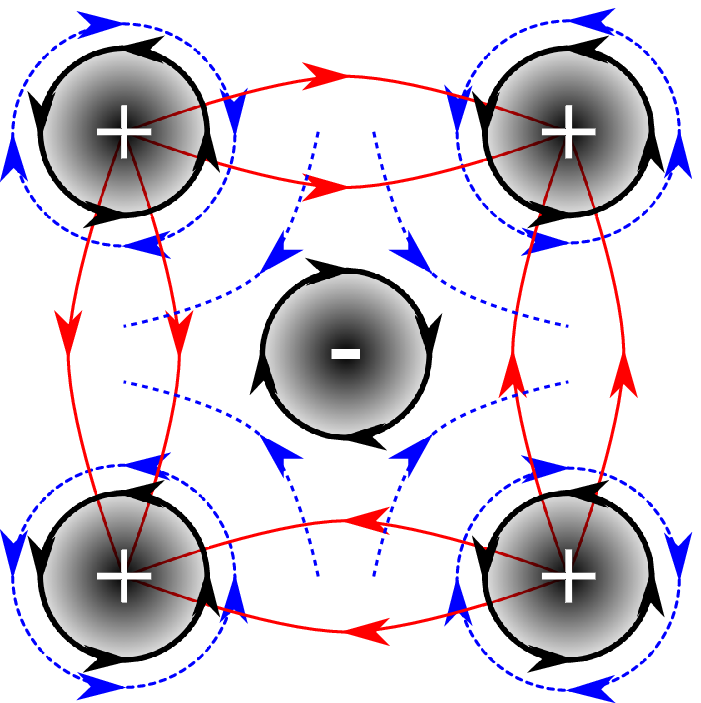}}
\caption{\label{HelicalLattice}Examples of helical spin-current vortex lattices: (a) the likely candidate state, (b) an alternative unstable state that costs more energy. The shaded circles are vortices (+) or anti-vortices (-) of $\theta$, with arrows showing the flow of ${\bf J}^z \sim \boldsymbol{\nabla}\theta$ currents. The solid red lines and arrows show the local orientation of $\hat{\boldsymbol{\theta}}$ (which is not a current flow), while the blue dashed lines and arrows show the local orientation of $\boldsymbol{\nabla}\alpha$ (which is related to the spin current $\vec{\bf J}$). Only a half of the unit-cell is shown. Note that each topological defect of $\theta$ is a ``coinciding vortex in $\eta_+$ and anti-vortex in $\eta_-$'' (or the other way round) according to (\ref{OP}), as stated in the {Fig.\VL} of the main article. The vortex lattice structure cannot be determined by the present analysis, and could be hexagonal as illustrated in the main article.}
\end{figure}

Any source of $\boldsymbol{\nabla}\alpha$ must be compensated by a nearby drain in order to avoid a buildup of currents that cost excessive $|\boldsymbol{\nabla}\alpha|^{2}$ kinetic energy. The $\boldsymbol{\nabla}\alpha$ source v.s. drain character is governed by the sign of $\hat{\boldsymbol{\theta}}\boldsymbol{\nabla}\theta$, which can be either plus or minus even when $n=1$ is fixed. Furthermore, every $n=1$ vortex of $\theta$ must be compensated by a nearby $n=-1$ anti-vortex in order to avoid the buildup of currents that would cost a forbiddingly high $|\boldsymbol{\nabla}\theta|^2$ kinetic energy. These considerations lead us to a candidate vortex lattice state shown in the Fig.\ref{HelicalLattice}(a). The $n=1$ vortices should form a bipartite lattice, whose one sublattice provides sources and the other drains for $\boldsymbol{\nabla}\alpha$. The plaquettes of this bipartite lattice should contain the $n=-1$ anti-vortices. It should be noted that the symmetry of this Abrikosov lattice naturally satisfies the conservation laws (\ref{CurCons5}) even near vortex cores.

A very similar alternative vortex lattice, shown in Fig.\ref{HelicalLattice}(b), could be envisioned with $\alpha$ having a vortex configuration just like $\theta$ (the average $\alpha$ ``vortex-charge'' density must also vanish, as required by the kinetic energy). This would imply $\nabla^{2}\alpha=0$, and the existance of sources and drains in the vector field of $\hat{\boldsymbol{\theta}}$. However, such a state is not stable because it violates the last current conservation equation of (\ref{CurCons6}). A partial remedy could perhaps be found according to the more accurate last equation of (\ref{CurCons5}), but only near the (-) anti-vortices. Related to this problem is a higher energy cost of this state (b) than the vortex lattice (a). Even though both types of vortex lattices keep $\hat{\boldsymbol{\theta}} \parallel (\hat{\bf z}\times \boldsymbol{\nabla}\alpha)$, the $\alpha$-vortices in the state (b) should have a large vortex ``charge'' in order to achieve the optimum values of $|\boldsymbol{\nabla}\alpha|$ in the bulk (recall that we are considering the limit of large spin-orbit couplings). This is bound to produce much larger and costlier vortex cores than the ones of the state (a), despite the fact that the $\theta$-vortices of both states always have the $n=\pm 1$ vortex ``charge''.

The geometry and other detailed properties of the helical triplet vortex lattice can be determined only numerically. This goes beyond the scope of the present analysis. Here we are content with knowing that the equilibrium superconductor of the Rashba-coupled Cooper pair triplets is a peculiar vortex lattice of helical spin-currents. This state is periodic as each vortex is matched by an anti-vortex in the unit-cell. There are more than one topological defects in a unit-cell due to the order parameter periodicity and other restrictions that we discussed. Consequently, these vortex lattices are likely more susceptible to quantum melting than the simplest hexagonal Abrikosov lattice of Cooper pairs in magnetic fields. Nevertheless, vortices and anti-vortices are kept firmly apart by the strong spin-orbit coupling. Their annihilation would result with the complete loss of helical spin-currents (in equilibrium), and thus cost a finite energy-density.

\section{Incompressible quantum liquids of helical triplets}

A vortex lattice of any kind can be melted by quantum fluctuations. The melting transition is generally first order since the vortex density is roughly constant. Namely, the separation between topological defects is set by the external gauge field that stabilizes a vortex state, and not by the overall strength of the superconducting order parameter that is being gradually reduced on the approach to the transition. The ensuing quantum vortex liquid is an incompressible state. If interactions fail to localize particles because their density is too low, this quantum liquid will generally exhibit many-body quantum entanglement and quasiparticles that have fractional statistics and quantum numbers.

The only theoretical approach capable of justifying the last extremely general claim is topological field theory. A theory of this kind has been constructed recently to describe both conventional and topological states of arbitrary spinor fields coupled to external gauge fluxes of arbitrary symmetry group and representation (see Ref. {\TFT} in the main article). Applied to the triplet spinor fields (\ref{OP}), this theory would have the imaginary-time Lagrangian density $\mathcal{L} = \mathcal{L}_\textrm{\pp LG} + \mathcal{L}_{\textrm{\pp t}}$:
\begin{eqnarray}\label{TopLGp}
&& \!\!\!\!\!\!\! \mathcal{L}_\textrm{\pp LG} = \eta^\dagger (\partial_0 - i\mathcal{A}_0) \eta
   + \Bigl\lbrack(\partial_{i}-i\mathcal{A}_{i})\eta\Bigr\rbrack^{\dagger} \frac{\widetilde{K}}{2}
     \Bigl\lbrack(\partial_{i}-i\mathcal{A}_{i})\eta\Bigr\rbrack \nonumber \\
&& ~~~~~  -\eta^{\dagger}\widetilde{t}\,\eta+\widetilde{u}|\eta|^4+\widetilde{v}|\eta^{\dagger}\gamma^{a}\eta|^2 \nonumber \\[0.1in]
&& \!\!\!\!\!\!\! \mathcal{L}_{\textrm{\pp t}} = \frac{i}{8}\, \eta^{\dagger}\epsilon^{\mu\nu\lambda}\Bigl\lbrack
   \partial_\mu^{\phantom{i}} \Bigl\lbrace\partial_\nu^{\phantom{i}},\Theta_{0}^{-1}\Bigr\rbrace \partial_\lambda^{\phantom{i}}
  +\Bigl\lbrace\partial_\mu^{\phantom{i}}\partial_\nu^{\phantom{i}}\partial_\lambda^{\phantom{i}},\Theta_{0}^{-1}
     \Bigr\rbrace\Bigr\rbrack\eta \ . \nonumber
\end{eqnarray}
The Landau-Ginzburg part $\mathcal{L}_\textrm{\pp LG}$ is essentially the same Lagrangian as that in (\ref{Act}), with only minor differences that make it more general. The topological term $\mathcal{L}_{\textrm{\pp t}}$ is completely inconsequential in any conventional phase, where either particles are localized (e.g. Mott insulators) or vortices are localized (e.g. superconductors). However, this term is allowed by symmetry and becomes important in any incompressible quantum liquid. Its role is then to determine the exchange statistics of excitations. It reduces to the well-known Chern-Simons Lagrangian of (fractional) quantum Hall states when the incompressible quantum liquids arise from Abelian external gauge fields $\mathcal{A}_\mu$.

The above topological field theory employs insight from the duality mapping to view incompressible quantum liquids as states where both particles and vortices are mobile but not condensed. Duality allows this only by ``attaching'' particles to vortices. It was argued in the Ref. {\TFT} that the incompressible quantum liquids shaped by the Rashba spin-orbit coupling are naturally non-Abelian, since the vortex bundle attached to a particle carries non-trivial helical spin-currents.

Here we only wish to summarize some expected features of the Rashba-based fractional TIs. Given that the parent superconducting state is a lattice of vortices and anti-vortices, the related vortex liquid state is fundamentally different from any quantum (spin) Hall state. The Rashba spin-orbit coupling violates spin-conservation and thus eliminates the quantization of the spin-Hall response. A bosonic quantum spin-Hall state is a vortex liquid with a finite density of vortex ``charge'', which we know how to microscopically describe by the Laughlin and other wavefunctions. In contrast, the incompressible quantum liquids we expect to obtain in our system have zero net density of vortex ``charge''. Note that its vortices and anti-vortices are kept apart against annihilation despite being in the liquid state, in order to retain fluctuating spin-currents that lower the Rashba energy. However, virtual pair annihilation and creation is possible. This does not jeopardize an elementary fractional vortex (quasiparticle) excitation in this liquid, because its excess vortex ``charge'' is conserved. The nature of vortex quasiparticles is hinted by the microscopic picture of vortices in the parent superconducting state, which we obtained in this supplement. It is currently not known what quantum numbers are carried by such excitations (other than charge), and how to construct an appropriate microscopic many-body wavefunction.

\end{document}